\documentstyle[prl,aps]{revtex}

\begin{document}

\twocolumn[\hsize\textwidth\columnwidth\hsize\csname  @twocolumnfalse\endcsname \title{ 
Vibration-induced jamming transition in granular 
media}
\author{G. D'Anna and G. Gremaud}
\address{Institut de G\'{e}nie Atomique, 
D\'{e}partement de Physique, Ecole Polytechnique F\'{e}d\'{e}rale de
Lausanne, CH-1015 Lausanne, Switzerland}
\date{\today}
\maketitle

\begin{abstract}The quasi-static frequency response of a granular medium
 is measured by a forced torsion oscillator method, with
 forcing frequency $f_{p}$ in the range $10^{-4}$~Hz to 5~Hz,
 while weak vibrations at high-frequency $f_{s}$, in the range
 50~Hz to 200~Hz,
 are generated by an external shaker.
 The intensity of vibration, $\Gamma $, is below the fluidization limit.
 A loss factor peak is observed in the oscillator response
 as a function of $\Gamma $ or $f_{p}$.
 In a plot of $\ln f_{p}$ against $1/\Gamma $,
 the position of the peak follows an
 Arrhenius-like behaviour over four orders of magnitude in  $f_{p}$.
 The data can be described as a stochastic hopping process
 involving a probability factor $\exp(-\Gamma _{j}/\Gamma )$
 with $\Gamma_{j}$ a $f_{s}$-dependent characteristic vibration intensity.
 A $f_{s}$-independent description is given by
 $\exp(-\tau _{j}/\tau )$,
 with $\tau_{j}$ an intrinsic characteristic time, and
 $\tau =\Gamma ^{n}/2\pi f_{s}$, n=0.5-0.6, 
 an empirical control parameter with unit of time.
 $\tau$ is seen as the effective average time during which the
 perturbed grains
 can undergo structural rearrangement.
 The loss factor peak appears as a crossover in the dynamic behaviour
 of the vibrated granular system,
 which, at the time-scale $ 1/f_{p}$, is solid-like at low $\Gamma$,
 and the oscillator is
 jammed into the granular
 material, and is fluid-like at high $\Gamma$, where the oscillator
 can slide viscously.
\end{abstract}
\vspace{-5pt}
\pacs{PACS numbers: 45.70.-n, 62.20.Dc, 81.05.Rm, 83.70.Fn}
\vspace{-25pt}
\vskip2pc]

\section{Introduction}

In a granular medium at rest the grains can be disposed in an enormous
number of different configurations. A weak external disturbance, but
powerful enough to overcome locally the friction force between two grains,
allows the granular systems to rearrange and to switch between these
''blocked'' configurations. The macroscopic behavior of a weakly disturbed
granular medium is, therefore, essentially controlled by statistical
properties of such transitions. It is of great interest to study this kind
of problem, since it may be a prototype of slow dynamics behaviors observed
in other physical systems\cite{Barrat2000}. The slow dynamics of weakly
disturbed granular media has been evidenced by the classical compaction
experiments of Knight {\it et al.}\cite{07 Knight} Another approach is the
experiment of Albert {\it et~al}.\cite{03 Albert}, in which a large solid
object is pulled slowly through a granular medium. The motion of the object
appears as resisted by chains of jammed particles\cite{01 Cates}\cite{02 Liu}%
, which support compressive stress. Beyond an elastic regime at very small
pulling force, the macroscopic motion of the object is a succession of
stick-slip events, where compressive stress is continuously built up in
particle chains, and abruptly released. At these successive unjamming events
the system switches between blocked configurations.

In this paper we implement a novel experimental method to study the problem:
we exploit a forced torsion oscillator\cite{04 DAnna oscillator} immersed in the
granular medium, and {\it in presence} of external weak vibration, as shown in
Fig. 1. We use a {\it dynamic method}, which provides more
information than the simple increase toward the unjamming threshold. In fact,
the amplitude of the angular displacement of the oscillator increases sharply
when the unjamming threshold forcing torque amplitude is approached, and at
the same time the angular displacement lags behind the sinusoidal torque.
This ''phase lag'' is determined by an energy dissipation which occurs in
the granular medium when grains start to slip one against the other. The
dynamic method gives access to both elastic and dissipation parameters of
the granular material during the slow dynamics.

\section{Experimental}

In the experiment, we hold a granular medium at a given high-frequency
vibration intensity, quantified by the normalized acceleration $\Gamma
=a_{s}\omega _{s}^{2}/g$, with $a_{s}$ and $f_{s}=\omega _{s}/2\pi $ the
amplitude and frequency of the vertical sinusoidal vibration, $g$ the
acceleration of gravity. At the same time, we measure the complex frequency
response, $G$, of the granular
medium (or the susceptibility $\chi=G^{-1}$)
 by a low-frequency forced torsion oscillator\cite{04 DAnna oscillator},
at the forcing frequency $f_{p}=\omega _{p}/2\pi $, with $f_{p}\ll f_{s}$.
In the oscillator method (see Fig. 1), the rotating probe of the oscillator is
immersed at a depth $L$ into a large metallic bucket (height 96 mm, diameter
94 mm) filled with glass beads of diameter $d=1.1\pm 0.05$~mm with smoothly
polished surfaces. The probe is covered by a layer of beads, glued on by an
epoxy, and its effective radius is $R_{e}$. All data presented here are
obtained with $L=20$~mm and $R_{e}\approx 2$~mm. We perform dynamic
experiments: the oscillator is forced into torsion oscillation by a torque $%
T\left( t\right) =T_{o}\exp (i\omega _{p}t)$ of frequency in the range $%
10^{-4}$~Hz to 5~Hz, and the angular displacement, $\theta (t)$, is
optically detected. An analyzer measures the complex frequency response of
the oscillator, given by $G=T/\theta$.
Typically we record the argument, $\arg (G_{1})$, and the absolute modulus, $%
|G_{1}|$, of the first harmonic, as a function of either $T_{0}$, $\Gamma $,
or $f_{p}$. We report the quantity $\tan [\arg (G_{1})]$, which for a linear
system coincides with the loss factor. The oscillator, when not immersed, can
be assumed elastic, with $T=G_{p}\theta $, where $G_{p}=18\times 10^{-3}$%
~N~m/rad is the torsion constant of the suspension wires. Notice that in
this work the maximum displacement of a point at the surface of the $2$~mm
probe is of the order of $0.1$~mm, i.e., much smaller than the glass bead
diameter.

An accelerometer provides a precise measurement of $\Gamma $, which can be
varied from $2\times 10^{-3}$ to above 1. The minimum value of $\Gamma $ is
limited by the accelerometer sensitivity. We vary $\Gamma $ by changing $%
a_{s}$ at fixed $f_{s}$, while $f_{s}$ is selected in the range 50~Hz to
200~Hz. The whole system is placed on an anti-seismic table.
Moisture-induced ageing effects\cite{07 Bocquet}\cite{04 DAnna oscillator},
and interstitial gas effects\cite{Pak}, are not observed for the large bead
size used here, and measurements are performed at uncontrolled ambient air.
In order to control compaction effects\cite{07 Knight}, all measurements are
taken in the same conditions, e.g., starting from a granular material shaken
at high $\Gamma $ and low $f_{s}$ for several minutes. Compaction effects
are apparently negligible in the time scale of the experiments for $f_{p}>0.01$%
~Hz, but may be present in the data at very-low frequency.

\section{Results}

A typical experimental result is reported in Fig. 2, which shows $\tan [\arg
(G_{1})]$ and $|G_{1}|$, measured as a function of the amplitude of the
applied torque $T_{0}$, for different $\Gamma $. With the vibrator {\it off}%
, that is for $\Gamma <2\times 10^{-3}$, the response is similar to the one
reported previously\cite{04 DAnna oscillator}, with a typical loss factor peak
at a torque denoted $T_{0}^{*}$ and a modulus step between two levels
denoted $G_{jam}$ and $G_{p}$. The dependence of the loss peak on the
geometrical parameters of the experiment is summarized\cite{04 DAnna
oscillator} by the empirical relation $T_{0}^{*}\propto \mu _{s}L^{2}R_{e}^{2}$%
, with $\mu _{s}$ the coefficient of static friction between the glass
beads. In this ''zero temperature-like'' conditions, the loss factor peak
can be easily explained: at very low applied torque, $T_{0}\ll T_{0}^{*}$,
the oscillator probe is jammed into the granular material, and only elastic
deformations arise, resulting in a purely elastic dynamic response, with a
negligible loss factor and a constant absolute modulus $G_{jam}$. By
increasing the torque amplitude, the oscillator probe unjams as the local
force between a pair of glass beads somewhere in the medium becomes large
enough for the two beads to slip one against the other, dissipating energy
by solid friction. The maximum ratio of dissipated over furnished energy,
that is a maximum of the loss factor, arises at $T_{0}^{*}$, which can be
seen as the average torque at which the oscillator probe unjams. At high
torque amplitude, $T_{0}\gg T_{0}^{*}$, the oscillator slides almost freely
into the granular medium, and the absolute modulus tends to the torsion
constant of the suspension wires, $G_{p}$.

With the vibrator {\it on}, one expects that the external vibration
facilitates the unjamming of the oscillator. By increasing $\Gamma $, the
modulus $|G_{1}|$ decreases monotonically, while $\tan [\arg (G_{1})]$ first
increases and then decreases, going through a maximum, as clearly visible in
Fig. 2. The fact that the external vibration drives the system through a
maximum in the loss factor is evidence that the vibration-induced
fluctuations can unjam the oscillator probe. Moreover, below $T_{0}^{*}$ the
response is essentially independent of $T_{0}$ , i.e., there is a linear
regime. The linearity is confirmed also by a negligible high harmonics
signal (not shown) for all $\Gamma $.

The general behavior in the linear regime is better rendered by Fig. 3 which
shows the previous maximum in the loss factor as characteristic ``jamming''
peaks observed as a function of $\Gamma $ for various forcing frequencies $%
f_{p}$. In Fig. 3 we obtain that for the low-torque amplitude, selected in
the linear regime, i.e., $T_{0}\ll T_{0}^{*}$, and at low $\Gamma ,$ the
applied torque alone is unable to unjam the oscillator probe and the response
is elastic. However, by increasing $\Gamma $, unjamming is induced by the
external vibration, and the response displays a loss factor peak. The data
shown in Fig. 3 are collected by decreasing $\Gamma $, but no difference is
observed in following runs if $\Gamma $ is successively increased, decreased
and so on, as shown in Fig. 4 for one of the curves of Fig. 3. We say that
the response is ''reversible'', although at a mesoscopic level energy is
continuously dissipated. The
loss factor peak can be seen as the crossover between two different
behaviours in the dynamics of the vibrated granular system:
at the time-scale set on by the forced oscillator, i.e., $ 1/f_{p}$,
the granular system appears solid-like at low-$\Gamma $, while it
appears fluid-like at high-$\Gamma $. It is a kind of
glass, or
jamming transition\cite{NicodemiJamT}{\it \ }, where the oscillator
gets stuck in the glassy granular medium.

Of course, since the ''jamming'' peak in Fig. 3 shifts with $f_{s},$ the
same peak can be observed as a function of the forcing frequency $f_{p}
$, as shown in Fig. 5, for various $\Gamma $. At high forcing frequency (but
still $f_{p}\ll f_{s}),$ the applied torque alone is unable to unjam the
oscillator probe and the response is elastic, with negligible loss factor and
modulus $G_{jam}.$ However, by decreasing $f_{p},$ i.e., by increasing the
time scale of the probing oscillator, the response evolves toward the one of
the unjammed oscillator, with modulus $G_{p}.$

From the shift of the previous ``jamming'' peaks with $f_{p}$ or $\Gamma $,
an Arrhenius-like semilogarithmic plot can be obtained, as shown in Fig. 6.
The data points, for a given $f_{s}$, obey an exponential behavior over four
decades in frequency. This is strong evidence for the underlying unjamming
process to be a statistical, activated-like hopping process and that some of
the usual statistical concepts of thermal systems can be extended to a
vibrated granular material. A first, obvious approach consists of formally
writing the rate of the hopping process as $R=\nu _{0}\exp (-\Gamma
_{j}/\Gamma )$, with $\Gamma _{j}$ a characteristic normalized acceleration
at which unjamming occurs, $\nu _{0}$ an attempt frequency, and $\Gamma $
the vibration intensity, playing the role of a temperature-like parameter. In
the linear regime, a peak in the loss factor is expected to arise when the
forcing frequency matches the hopping rate, i.e., when $\omega _{p}=R$, and $%
\Gamma _{j}$ appears as the ''slope'' of a straight line in a plot of $\ln R$
against $1/\Gamma $; however, then $\Gamma _{j}$ depends on $f_{s}$ (see
Fig. 6), which means that $\Gamma _{j}$ is not an intrinsic property of
the granular material.

To overcome this difficulty, we search a ''scaling'' of $\Gamma $ and $f_{s}$
which eliminates the $f_{s}$ dependence. We find that as a function of
the inverse of $\Gamma ^{n}/\omega _{s}$, with $n=0.5$, the data for different $%
f_{s}$ have almost the same ''slope'', as shown in Fig. 7. Hence, we
write the hopping probability per unit time as $R=\nu _{0}\exp (-\tau
_{j}/\tau )$, with $\tau _{j}$ a characteristic time of the unjamming
process, and $\tau =\Gamma ^{1/2}/\omega _{s}$ a parameter
which has the unit of time\cite{notaGamma}. (Fitting values are given in
Fig. 7) Alternatively (Fig. 8), we find also that a ''scaling'' of the form $\tau =\Gamma ^{n}/\omega _{s}$, with $n=0.57$, almost collapses the data of Fig. 6 over the same straight line.
This empirical definition
is appealing since both the parameters $\tau $ and $\tau _{j}$ we introduce,
are independent of the vibration frequency $f_{s}$. The unjamming time $\tau
_{j}$ experimentally is an {\it intrinsic} parameter of the granular system,
which for a given $L$ and $R_{e}$ is likely to depend on ''mesoscopic''
parameters such as the grain size and shape, and on ''microscopic''
parameters controlling the nature of the contact forces between grains. For $n=0.5$ the
externally controlled parameter $\tau $ appears as a typical time of the
vibrated granular system in the gravitational field: $\Gamma ^{1/2}/\omega
_{s}=(a_{s}/g)^{1/2}$ is the time of flight of a body, initially at rest,
falling for a distance $a_{s}$. (It is also the period of a simple oscillator
with thread $a_{s}$ and freely swinging.) Notice that $\tau $ is not a
kinetic energy-type temperature, as defined for vigorously vibrated gas-like
granular phases; indeed, for weakly excited granular systems,
configurational statistics on slow degrees of freedom can be decoupled from
fast kinetic aspects\cite{Barrat2000}\cite{Edwards}\cite{Clement}\cite
{Cugliandolo2000}\cite{Nicodemi}\cite{Mehta2000}.

\section{Discussion}

That the rate of our ''activated-like'' process is better given by
a probability factor involving the ratio of characteristic times, $\exp
(-\tau _{j}/\tau )$, and not, e.g., by the ratio of vibration intensities, $%
\exp (-\Gamma _{j}/\Gamma )$, is not surprising since the fundamental
phenomenon is the rate of energy dissipation. What is the mesoscopic nature
of the hopping processes at the length scale of the glass beads? Since we
observe an elastic regime at low $\Gamma $, we conclude that the oscillator
probe is completely jammed and no dissipative events (i.e., no slipping
events) arise. We can suppose the system to oscillate elastically around one
unique blocked configuration. In the picture, below the loss factor peak,
the external high-frequency vibration propagates in the system as elastic
fluctuations only. Such elastic fluctuations are non-dissipative vibrations
of the force-chain network which holds the oscillator probe in place. By
increasing $\Gamma $, a large number of increasingly energetic elastic
fluctuations arise (or increasingly {\it longer} if we focus on the
parameter $\tau =\Gamma ^{n}/\omega _{s}$), and the force-chain network
can be seen as exploring different elastic configurations until a critical
configuration is reached. A critical configuration is such that, somewhere,
the local friction force between two glass beads is overcome and slipping
arises, momentarily unjamming the oscillator probe. The system can switch to
another blocked configuration.

A single slipping event possibly triggers a large-scale non-elastic
rearrangement of the beads, that is an internal micro-avalanche involving
two or more beads. (We emphasize that a massive object moving
 into a granular medium can introduce local fluidization. The inertia
 acquired by the object during a slip may be large enough to overcome the resisting
 force of the granular medium, and the object moves further by
 successive failure, or "inertial" fluidisation, of the resisting grains arrangement. This effect can be reduced by immersing the object 
 deep enough, so that the resisting force is larger than the inertial force.)
 Afterwards, the oscillator probe gets jammed by a new
force-chain network in a slightly different position. The macroscopic slow
rotation of the oscillator is a sequence of stick-slips, where large
fluctuations causing a slip are rare events if compared to the numerous
elastic fluctuations. Hence, the dynamics is controlled by the extreme
fluctuations in the force-chain network, even if elastic, or under-critical
fluctuations occur in much larger number. Since a slipping event involves
inelastic microscopic processes at the interface between grains, such as
plastic and viscoplastic deformation, fatigue, surface fracture, blow out of
capillary bridges, and other forms of localized dissipative processes\cite
{Bowden}, a slipping event requires a {\it minimum finite time} to occur.
Let this time be $\tau _{j}$. The ''thermal'' time $\tau $ can be seen as
the {\it average} time-window during which the grains have some freedom to
rearrange their position and, possibly, reach a critical slipping
configuration and unjam. As a consequence, unjamming is determined by the
occurrence probability of a window time $\tau _{a}$ greater as, or equal to $%
\tau _{j}$. Even though we do not know the probability distribution of $\tau
_{a},$ according to extreme order statistics theory\cite{Galambos}\cite{9
Vinokur}\cite{11 Sollich}\cite{12 Bouchaud}, we can speculate the occurrence
probability of unjamming events to be $\exp (-\tau _{j}/\tau )$. This gives
the rate of the extreme fluctuations, namely $R=\nu _{0}\exp (-\tau
_{j}/\tau )$.

In summarizing, we observe a peak in the loss factor as a function of the
empirical control parameters $\tau $ or $\Gamma .$ The peak can be viewed as
a crossover, at the time-scale $ 1/f_{p}$ set on by the forced oscillator,
 in the dynamics of the vibrated granular medium:
such a crossover separates
a ''low-temperature'' (short $\tau ,$ or small $\Gamma $) solid-like behaviour
where the oscillator probe is jammed in the granular medium, from a
''high-temperature'' (long $\tau ,$ or large $\Gamma $)
 fluid-like dynamic behaviour.
 This crossover follows an Arrhenius-like form in the $\ln f$
vs. $1/\Gamma $ plane,
 reminiscent of the
mechanical response of usual glass-froming materials.

Figure captions

FIG. 1. Sketch of the forced torsion oscillator immersed at a depth $L$ into a
granular medium of glass beads. A single layer of glass beads is glued to
the oscillator probe, and the effective radius is $R_{e}$. The container,
filled with the granular material, is shaken vertically by a vibrator at the
intensity of vibration, $\Gamma $, below the fluidization limit. The method
provides a measure of the complex frequency response of the granular medium
while the vibrator mimics ''thermal'' fluctuations. 1=suspension wires;
2=permanent magnet; 3=external coils; 4=mirror; 5=probe; 6=vibrator.

FIG. 2. Oscillator frequency response, at $f_{p}=1$~Hz, as a function of the
amplitude of the applied torque $T_{0}$, for different vibration intensities $%
\Gamma $, for $f_{s}=200$~Hz. (a) The loss factor, given by $\tan [\arg
(G_{1})]$, versus $T_{0}$. The position of the peak obtained at $\Gamma
<2\times 10^{-3}$ (i.e., with the vibration off) is denoted $T_{0}^{*}$. (b)
The modulus $|G_{1}|$ versus $T_{0}$. The two extreme levels on the curve
obtained at $\Gamma <2\times 10^{-5}$ are denoted $G_{p}$ and $G_{jam}$
respectively. For $T_{0}\ll T_{0}^{*}$, the response is independent of $%
T_{0} $ , i.e., there is a linear regime, confirmed also by a negligible
high harmonics signal (not shown) for all $\Gamma $.

FIG. 3. Oscillator frequency response as a function of the vibration intensity $%
\Gamma $, with $f_{s}=200$~Hz, for different forcing frequencies $f_{p}$ of
the oscillator, at a given low torque amplitude $T_{0}=3.2\times 10^{-5}$~N~m
selected in the linear regime, i.e., $T_{0}\ll T_{0}^{*}$. (a) The loss
factor $\tan [\arg (G_{1})]$, versus $\Gamma $. (b) The modulus $|G_{1}|$
versus $\Gamma $. For each $f_{p}$ a peak with a maximum at a vibration
intensity $\Gamma ^{*}$ can be seen. The data shown in Fig. 3 are collected
by decreasing $\Gamma $, but no difference is observed in following runs if $%
\Gamma $ is successively increased, decreased and so on (see Fig. 4.).

FIG. 4. Similar to Fig. 3, for $f_{p}=1$ Hz, but for $\Gamma $ successively
decreased, increased, decreased and so on.

FIG. 5. Oscillator frequency response as a function of the forcing frequency $%
f_{p}$, for different vibration intensity $\Gamma $, with $f_{s}=200$~Hz, at a
given torque amplitude $T_{0}=3.2\times 10^{-5}$~N~m. (a) The loss factor $%
\tan [\arg (G_{1})]$, versus $f_{p}$. (b) The modulus $|G_{1}|$ versus $%
f_{p} $. One can see for each $\Gamma $ a peak with a maximum at a frequency 
$f_{p}^{*}$. The data shown are collected by decreasing $f_{p}$. A Debye
peak of equation $C\omega _{p}\tau _{c}/(1+\omega _{p}^{2}\tau _{c}^{2})$
with $C=1.9$ and $\tau _{c}=R^{-1}=0.6$ is shown in (a). As a function of
the frequency, the shape of the Debye peak is independent of the exact
definition of the temperature-like parameter entering the rate $R$. The
observed ''jamming'' peaks are much larger than the pure Debye peak,
suggesting that the underlying dynamics is glassy in nature.

FIG. 6. The semilogarithmic Arrhenius-like plot reporting the forcing
frequency $f_{p}$ versus $1/\Gamma $ of ''jamming'' peaks similar to the
ones in Figs. 3 and 5 (filled symbols from measurements vs. $\Gamma $; open
symbols from measurements vs. $f_{p}$), for different vibration frequencies $%
f_{s}$. The data are fitted (dashed lines) by $2\pi f_{p}=\nu _{0}\exp
(-\Gamma _{j}/\Gamma )$, which gives $\nu _{0}\approx $70~Hz and $\Gamma
_{j}\approx $0.014 for $f_{s}=50$~Hz, $\nu _{0}\approx $66~Hz and $\Gamma
_{j}\approx $0.042 for $f_{s}=$100~Hz, and $\nu _{0}\approx $49~Hz and $%
\Gamma _{j}\approx $0.14 for $f_{s}=$200~Hz.

FIG. 7. The forcing frequency $f_{p}$ versus the inverse of the empirical
control parameter $1/\tau $, i.e., versus $\omega _{s}/\Gamma ^{n}$,
 with $n=1/2$.
 The data are fitted (plain and dashed lines) by $2\pi f_{p}=\nu _{0}\exp
(-\tau _{j}/\tau )$, which gives $\nu _{0}\approx $1336~Hz and $\tau
_{j}\approx 1.3\times 10^{-3}$~s for $f_{s}=50$~Hz, $\nu _{0}\approx $%
1465~Hz and $\tau _{j}\approx 1.2\times 10^{-3}$~s for $f_{s}=100$~Hz, $\nu
_{0}\approx $2698~Hz and $\tau _{j}\approx 1.2\times 10^{-3}$~s for $%
f_{s}=200$~Hz. The straight lines have almost the same ''slope'' $\tau _{j}$%
. The average is $\left\langle \tau _{j}\right\rangle =1.26\times 10^{-3}$%
~s. $\nu _{0}$ is seen as a natural vibration frequency of the granular
medium. Considering the present precision of the data, no clear relationship
between $\nu _{0}$ and $f_{s}$ can be found, even though $\nu _{0}$
increases as $f_{s}$ increases.

FIG. 8. Similar to Fig. 7, but with $n=0.57$. The data are fitted (plain and dashed lines) by $2\pi f_{p}=\nu _{0}\exp
(-\tau _{j}/\tau )$, which gives $\nu _{0}\approx $648~Hz and $\tau
_{j}\approx 8.0\times 10^{-4}$~s for $f_{s}=50$~Hz, $\nu _{0}\approx $%
683~Hz and $\tau _{j}\approx 7.8\times 10^{-4}$~s for $f_{s}=100$~Hz, $\nu
_{0}\approx $1019~Hz and $\tau _{j}\approx 8.7\times 10^{-4}$~s for $%
f_{s}=200$~Hz. 


\begin{references}
\bibitem{notaGamma}  When the vibration frequency is not a control parameter,
one can write the ``activated'' behavior in term of probability factors
involving $\Gamma $ instead than $\Gamma ^{n}/\omega _{s}.$ As shown in
Fig. 6, for a fixed vibration frequency $f_{s},$ formally writing the rate of
the activated-like hopping process as $R=\nu _{0}\exp (-\Gamma _{j}/\Gamma )$%
, suffices.
\end{references}
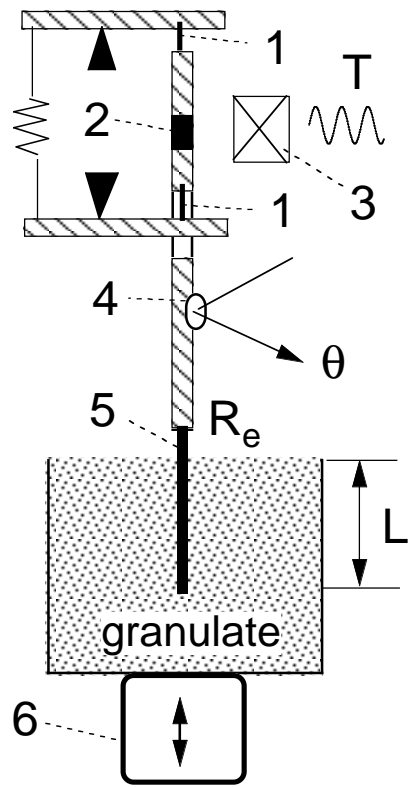
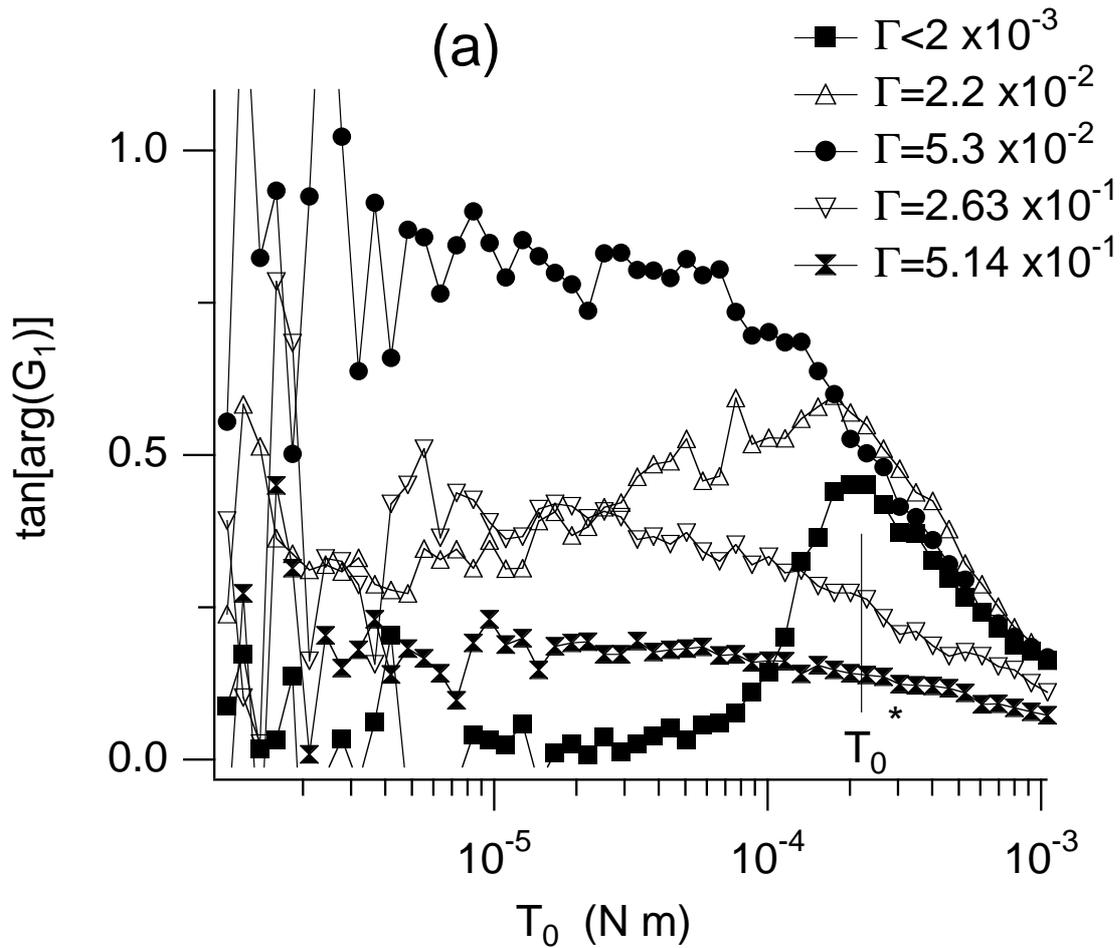
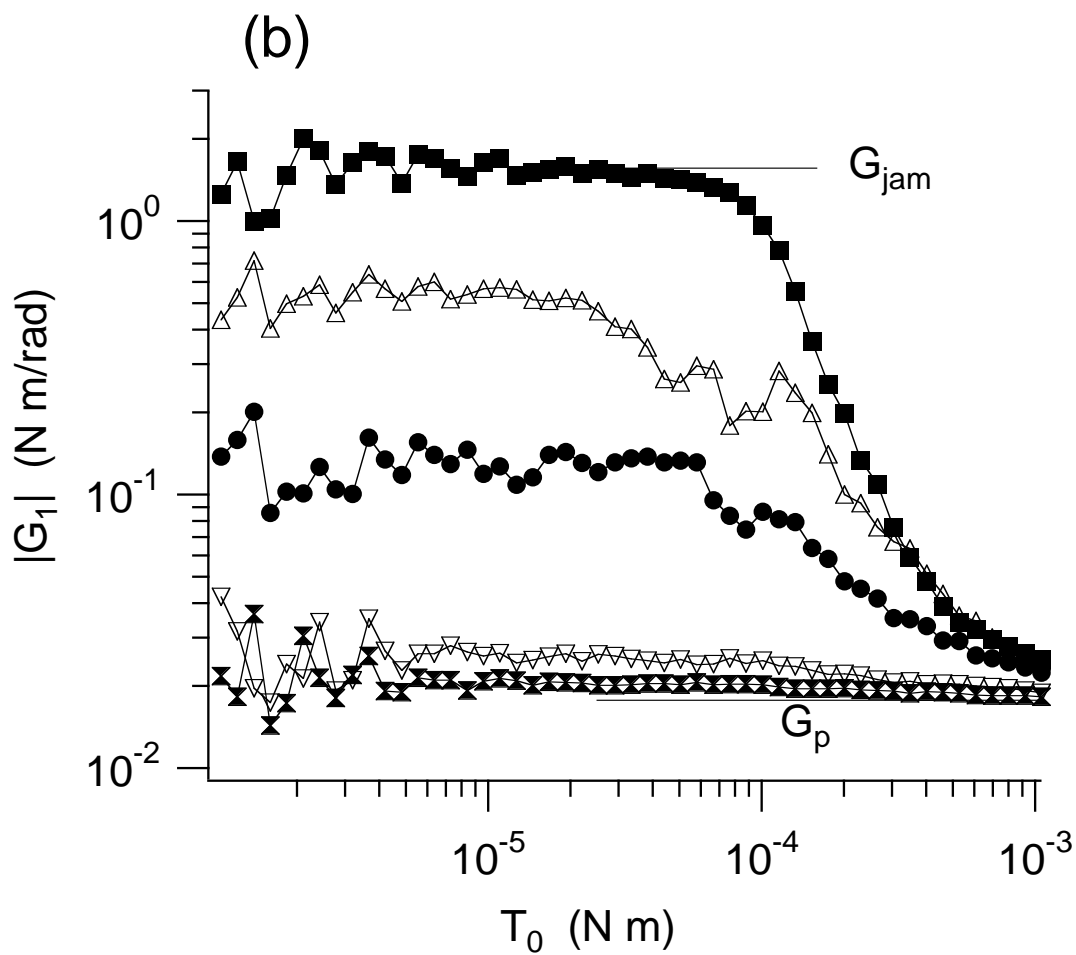
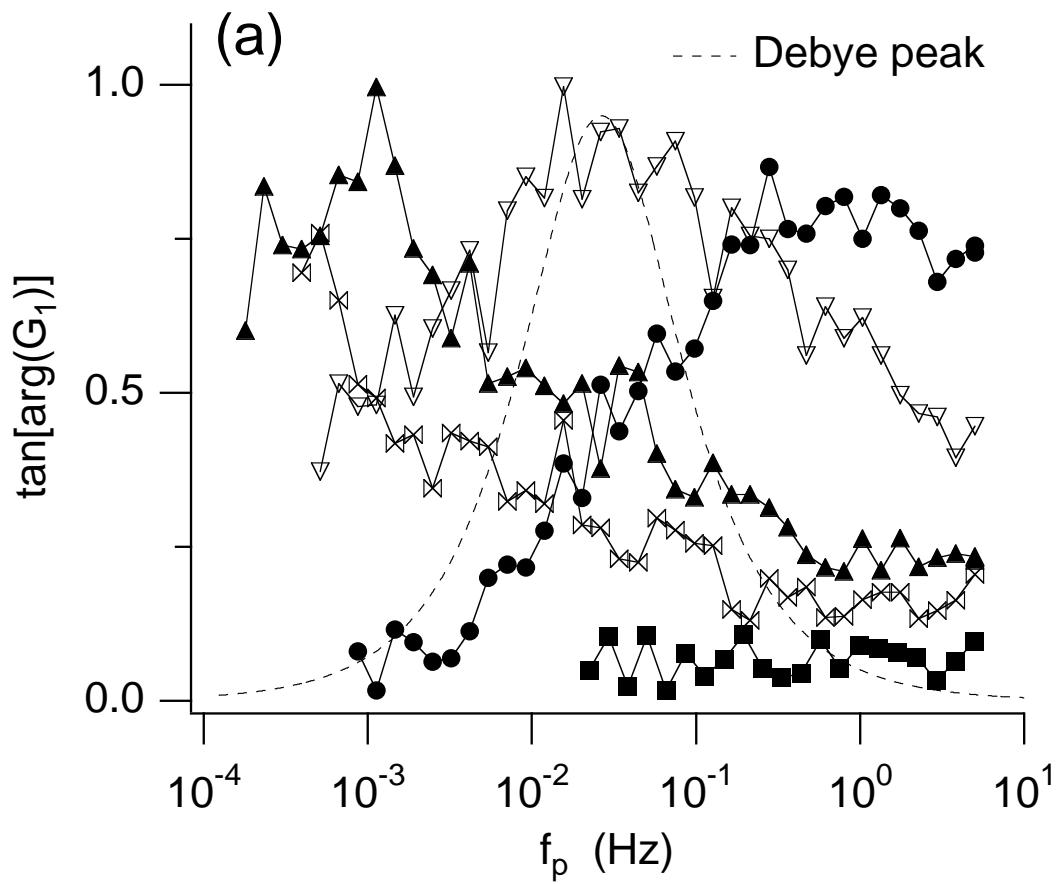
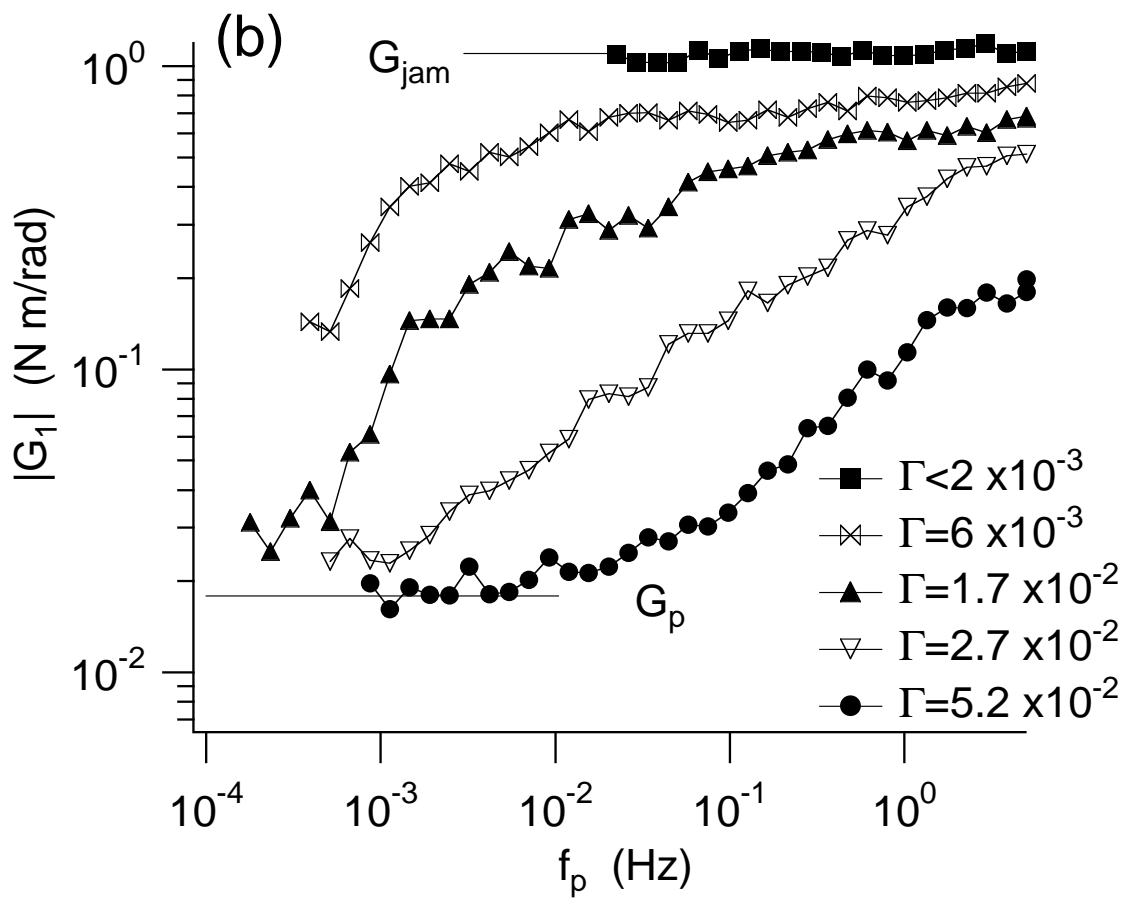
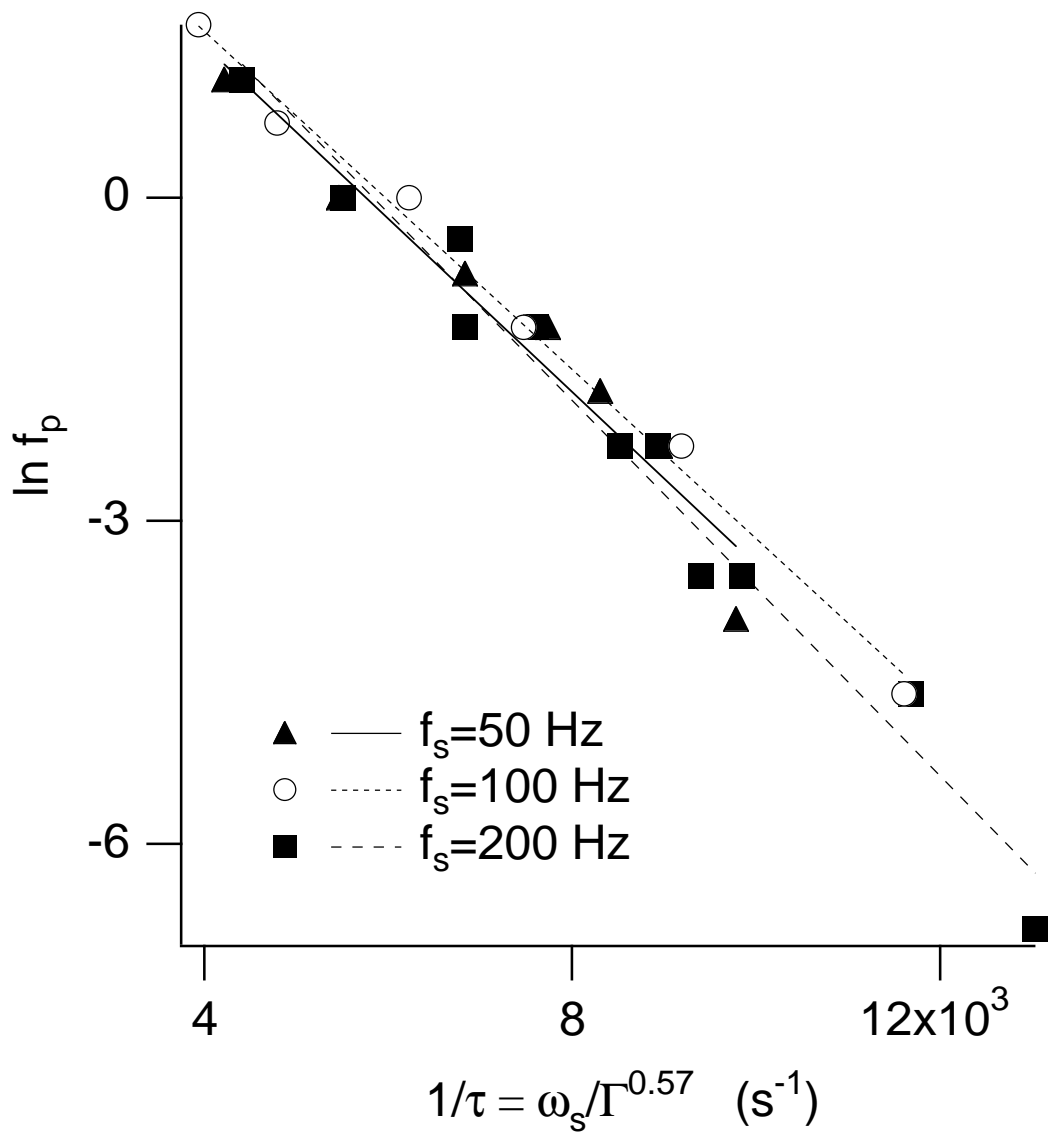
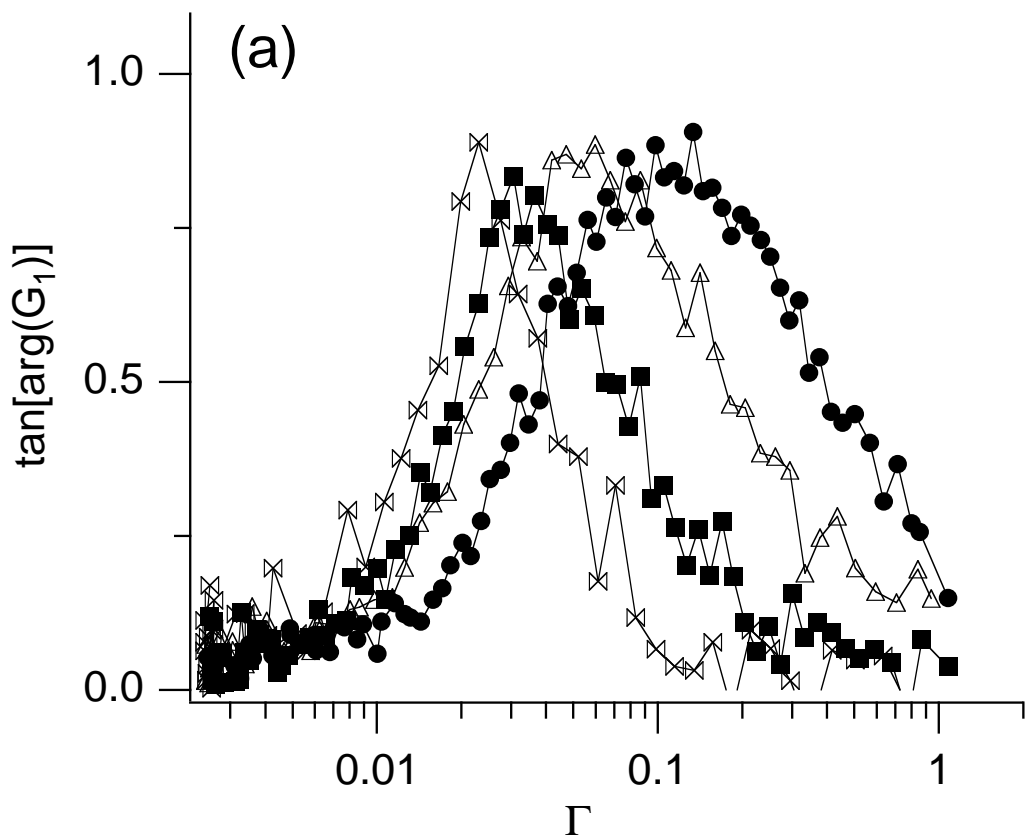
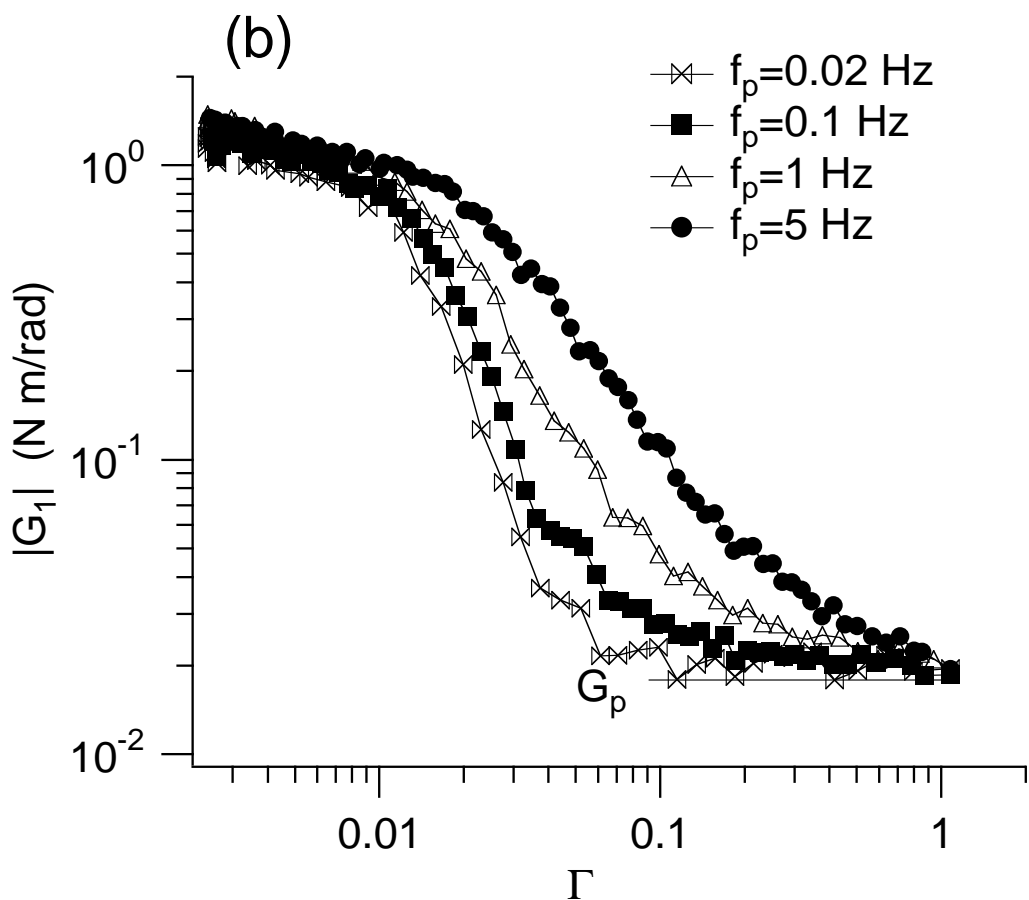
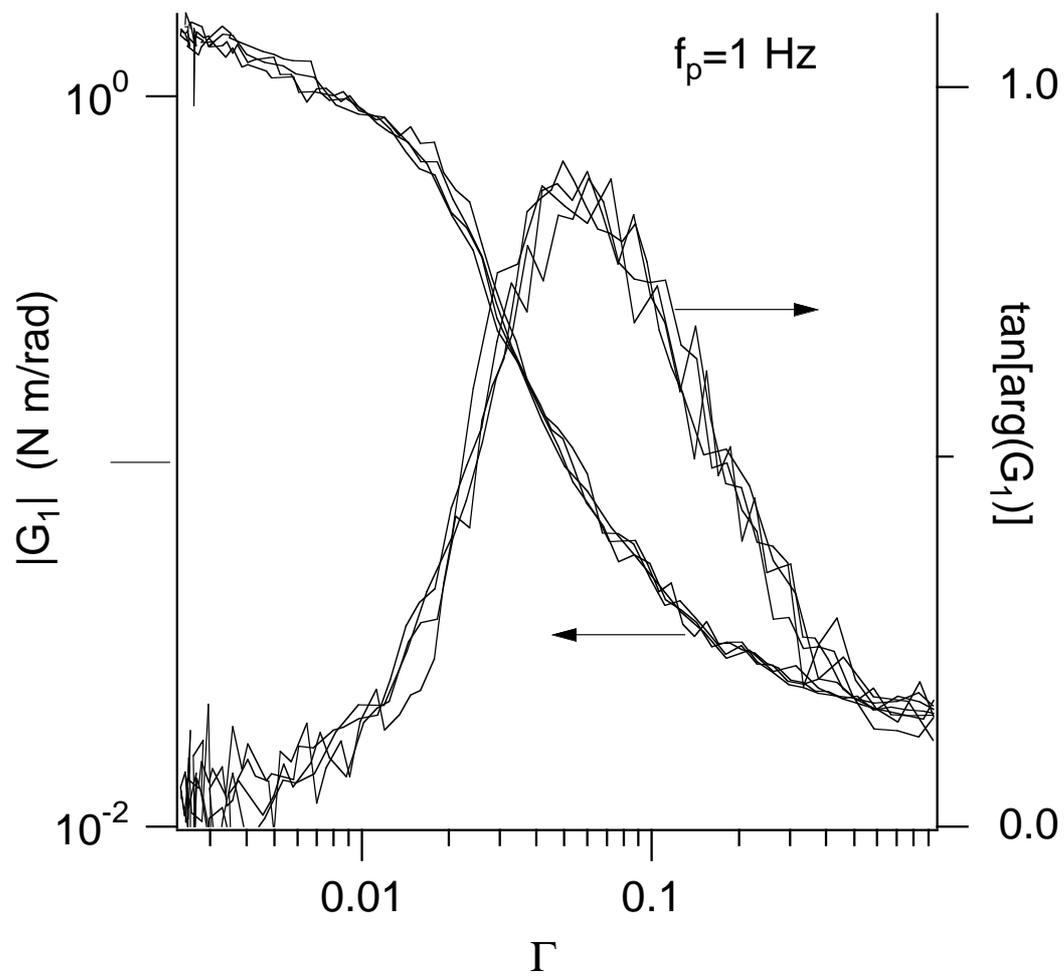
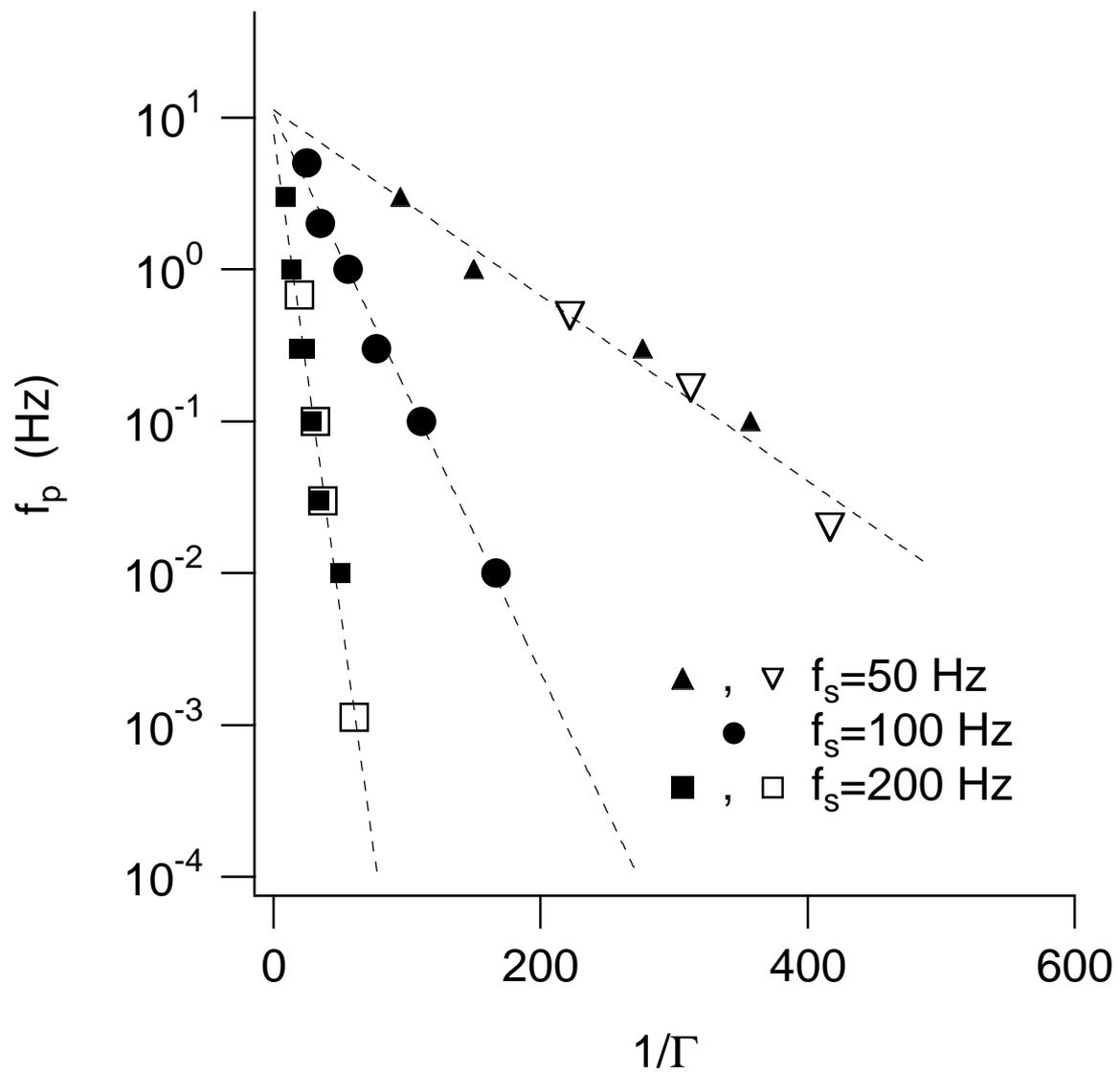
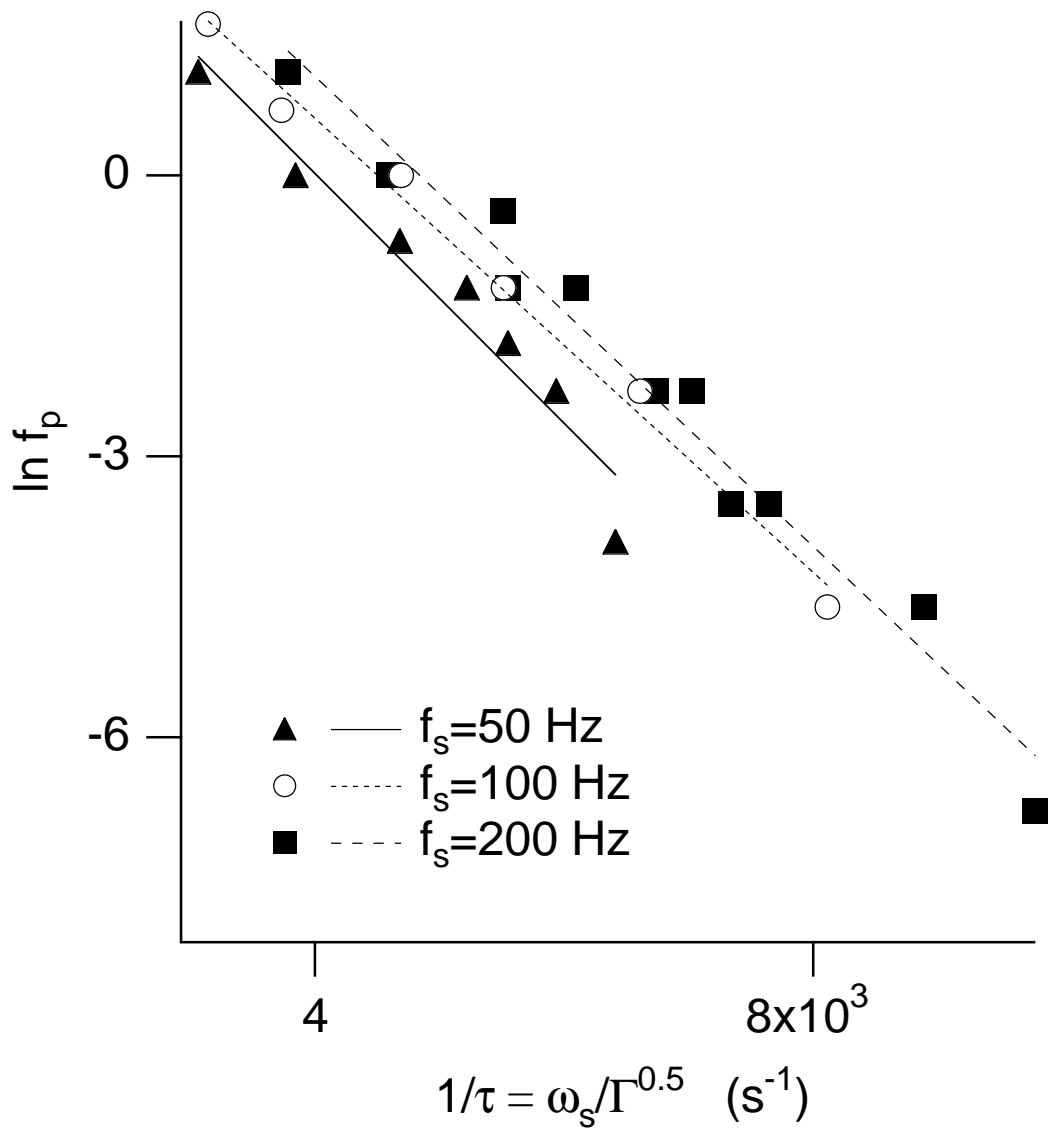
\end{document}